\documentstyle[aps,prd,epsfig]{revtex}

\newcommand{\Hu}{{\cal H}} \newcommand{\Ka}{{\cal K}}
 \newcommand{\cs}{{\Upsilon}}

\begin{document}
\tighten \draft \twocolumn[\hsize\textwidth\columnwidth\hsize\csname
@twocolumnfalse\endcsname

\author{Patrick Peter$^1$ and Nelson Pinto-Neto$^2$}

\address{$^1$Institut d'Astrophysique de Paris, 98bis boulevard Arago,
75014 Paris, France.\\ $^2$Centro Brasileiro de Pesquisas F\'\i sicas,
Rue Dr.  Xavier Sigaud 150, Urca 22290-180 -- Rio de Janeiro, RJ,
Brazil}

\title{Has the Universe always expanded~?}

\date{11 September 2001}

\maketitle

\begin{abstract}
We consider a cosmological setting for which the currently expanding
era is preceded by a contracting phase, that is, we assume the
Universe experienced at least one bounce. We show that scalar
hydrodynamic perturbations lead to a singular behavior of the Bardeen
potential and/or its derivatives (i.e. the curvature) for whatever
Universe model for which the last bounce epoch can be smoothly and
causally joined to the radiation dominated era. Such a Universe would
be filled with non-linear perturbations long before nucleosynthesis,
and would thus be incompatible with observations. We therefore
conclude that no observable bounce could possibly have taken place in
the early universe if Einstein gravity together with hydrodynamical
fluids is to describe its evolution, and hence, under these
conditions, that the Universe has always expanded.
\end{abstract}

\pacs{98.80.Hw, 98.80.Cq}

\narrowtext
\vspace{0.2cm} ]

\section{Introduction}

Since the development of singularity theorems~\cite{singularity}, it
has been a paradigm that the Standard Big Bang (SBB)
cosmology~\cite{SBB} requires the Universe to have emerged from an
initial singularity. However, as all theorems, those concerning
singularities were based on a set of hypothesis. In the 1970's and
after, many non-singular cosmological models with bounces were
constructed~\cite{seventies},\footnote{Already in the 1930's, a
non-singular cosmological model was proposed by Tolman~\cite{tolman}.}
where indeed one or more of the hypothesis of the theorems were
violated, such as, for instance, energy conditions or the validity of
Einstein gravity.

More recently, superstring-inspired models suggested that this
singularity might have been avoided if gravity was modified somehow in
the near Planckian regime for which a maximal curvature was supposed
to be reachable. Note that although this model seems to have always
expanded in the Jordan's frame, a bounce behavior is also present in
the Einstein frame. Such a pre Big Bang (PBB) scenario~\cite{PBB} is,
however, plagued with various problem, as, e.g. the graceful
exit~\cite{grace}, and fails to reproduce the primordial perturbation
spectrum as observed in the recent Cosmic Microwave Background (CMB)
data~\cite{CMBdata}. The non-singular hypothesis was revigorated even
more recently with the {\sl Ekpyrotic Universe} proposal~\cite{Ekp}.

Quantum gravity also provides new ways out of the singularity
theorems. For instance, models exist in which an expanding Universe
region might originate from a Black Hole
collapse~\cite{brandy}. Quantum cosmology, by the Wheeler-de Witt
equation, also yields bouncing behaviors for the scale
factor~\cite{bounce,bounce2}. However, in this case, whether the
Universe actually avoids the singularity may depend on the
interpretation that is given to its wave function. In particular, in
the many-world view~\cite{many}, it is the expectation value of the
scale factor that bounces, so that, in practice, nothing can be said
as to the value it takes in the realization corresponding to our
Universe.  However, in the ontological interpretation~\cite{onto} of
the wave function of the Universe, the Bohmian trajectory, being
meaningful, implies an actual physical bounce. (Note the interesting
fact that this interpretation also offers a natural
solution~\cite{graceBohm} to the graceful exit problem in the PBB
scenario.)

In this paper, we restrict our attention to pure general relativity
(GR) and examine a Friedman-Robertson-Walker (FRW) setting whose
stress-energy source comes from a collection of perfect fluids (having
vanishing anisotropic stresses).  We then assume that a bounce took
place at some instant in time and concentrate on observationally
meaningful models, namely those for which the bounce epoch can be
smoothly and causally connected to a radiation dominated phase where
Big Bang Nucleosynthesis (BBN), assumed as a physical requirement in
whatever model, can safely take place. Such Universes will be called
``realistic'' hereafter. What precedes the bounce itself is also left
out of our analysis which is thus not restricted to non-singular
models. In short, we just assume that, for some reason, our Universe
has not always expanded, and we question the observable consequences
of this possibility.

This work is to be contrasted with Ref.~\cite{Visser} in which the
mathematical conditions  on the total fluid for the  bounce to occur
are derived (see the Appendix for a physical restriction of these
conditions)~: our analysis is in some sense a follow-up of this work
as we examine stability.

In the framework defined above, we investigate the behavior of scalar
adiabatic perturbations near the bounce through the Bardeen
gauge-invariant potential. We show that these perturbations can grow
without limit in two situations. One, when the Universe passes through
the bounce itself. The other, if the existence of the bounce imposes a
violation of the Null Energy Condition (NEC) (and hence of all the
other energy conditions~\cite{Visser}), when the Universe makes the
transition from the region where the NEC is violated to the region
where it is not.\footnote{Fluids which violate the NEC are called
exotic matter. They have strange properties but there is no theorem
asserting their impossibility on macroscopic scales \cite{thorn} (one
way or the other) For examples of energy-momentum tensors where the
NEC is violated, see Ref.  \cite{nec}}.  In all cases of physical
interest, at least one of these instabilities occurs. These facts, in
turn, imply that there must exist a point in time where, apart for
infinitely fine-tuned initial conditions (and yet for all wavenumbers
at the same time), the perturbations grow unbounded. This means that
if such a bounce had occurred, then very large inhomogeneities, at all
scales, would have been formed very early on, that would presumably
have turn non linear very rapidly, i.e. long before
nucleosynthesis. Such a conclusion is in flagrant contradiction with
BBN, so that we are led to postulate a {\sl no-bounce
conjecture}~\cite{Ja-No-Bounce}.

In fact, our conclusion is more general than that. Even without
bounce, the point (in time) where the NEC is recovered (assuming it
has been violated in the past, which is the case for the majority of
bouncing Universes) yields divergences in the primordial
perturbations.  As a result, the PBB models, which may have no bounce,
acquires a new constraint as follows~:  the graceful exit problem, in
order to be solved, also requires, in many instances~\cite{veneziano},
a violation of the NEC for a finite amount of time (see however
Ref.~\cite{graceBohm}). Models should then take into account the fact
that the recovery of the NEC should not occur at a time where our
hypothesis (GR plus hydrodynamical perturbations) are already valid to
avoid the excess perturbation problem.

Finally, there is the special case, which is actually not a bounce,
for which the scale factor, although never decreasing in
time, happens to stop at an instant ($a(\eta _0) = a_0 + b \eta _0 ^{2i+1} + ...$
for some $\eta _0$, with $i \geq 1$). Such a ``pseudo-bounce'' situation
is included in our analysis, and it
also generates a cosmological catastrophe. One must therefore
conclude that the expansion must be {\it strictly} monotonous in order
to make sense.

In what follows, after a short review (Sec.~\ref{sec:rappels}) on the
essentials of cosmological perturbation theory in the framework with
which we are concerned, heavily based on Ref.~\cite{mfb}, we detail
the structure of a bouncing epoch (Sec.~\ref{sec:bounce}) by making an
expansion of the scale factor in powers of the conformal time. We then
go on to evaluate the evolution of the primordial perturbations (in a
gauge invariant manner in order to avoid spurious modes) around the
bounce in Sec.~\ref{sec:pert}, establishing our
results. Secs.~\ref{sec:discuss} and \ref{sec:conc} are devoted to
discussions and conclusions, while a final appendix shows that
connecting the bouncing phase with the observable Universe  yields
constraints on the parameters appearing in the expansion and exhibits
clearly the violation of the NEC in many cases.

\section{classical cosmological perturbation theory and the Null
Energy Condition}
\label{sec:rappels}

The purpose of this section is to recall various definitions and
evolution equations for scalar hydrodynamic perturbation, for the sake
of self-consistency.

Cosmological perturbations are expected to be produced quantum
mechanically during the early stage of the evolution of the Universe.
Expansion drives the quantum mechanical fluctuations of any field
present in one's favorite model to classical values, which then evolve
according to GR. Although we will concentrate on the subsequent
evolution of primordial perturbations as soon as they have been
produced, we shall also summarize the actual process of building them
as the quantum modes of the field.  This quantum analysis yield
different insights into the nature of the problem.

The background metric we shall be using is
\begin{equation} \hbox{d}s^2 = a^2(\eta) \left( \hbox{d}\eta^2 -
\gamma_{ij} \hbox{d}x^i \hbox{d}x^j \right), \end{equation} with
$\eta$ the conformal time, related to the cosmic time $t$ through the
scale factor $a$, $a \hbox{d}\eta = \hbox{d}t$, and
\begin{equation} \gamma_{ij} \equiv {\delta_{ij}\over [1 +{1\over 4}
\Ka {\bf x}]^2} \end{equation} the 3-space metric. Here, $\Ka =0,\pm
1$ is the sign of the curvature. The general form of metric
perturbations on this background reads
\begin{eqnarray} \hbox{d}s^2 &=& a^2(\eta) \left\{ (1+2 \phi
)\hbox{d}\eta^2 - 2 B_{;i}\hbox{d}\eta\hbox{d} x^i \right. \nonumber
\\ & & - \left. \left[ (1-2\psi)\gamma_{ij} + 2 E_{;ij}\right]
\hbox{d}x^i \hbox{d}x^j \right\}\; .\label{dg}\end{eqnarray}  {}From
this form, it is convenient to define the gauge-invariant Bardeen
potentials~\cite{bardeen}
\begin{equation} \Phi = \phi + {1\over a} [(B-E')a]',\end{equation}
\begin{equation} \Psi = \psi - {a'\over a} (B-E'),\end{equation}
(where a prime means a derivative with respect to conformal time), in
terms of which it is possible to expand the perturbed Einstein
equation in a gauge independent way. The latter are further simplified
in the particular case of vanishing anisotropic stress for which $\Psi
= \Phi$. In what follows, we shall restrict our attention to this
special case. Note that for perturbation theory to be valid, and space
not to become very inhomogeneous, the requirement $|\Phi|\ll 1$ should
hold at all times and at all wavelengths of perturbation. The stress
energy tensor, source of Einstein field equations, will take the form
\begin{equation} T^\mu_\nu = (\epsilon + p)u^\mu u_\nu - p
\delta^\mu_\nu, \label{Tmunu}\end{equation} for energy density
$\epsilon$, pressure $p$ and 4-velocity {\bf u}. Following
Ref.~\cite{singularity}, one then defines the Null Energy Condition
through the Null Convergence Condition which requires the Ricci tensor
to satisfy $R_{\mu\nu} n^\mu n^\nu \geq 0$ for all null vector
$n^\mu$. If this condition is fulfilled, then Einstein equations imply
that, with the form~(\ref{Tmunu}) of the stress-energy tensor, the
relation
\begin{equation} \hbox{NEC}\Longleftrightarrow \epsilon + p \geq
0,\label{NEC}\end{equation} holds. This relation, whose violation
implies the violation of all other energy conditions, is indeed
assumed to hold in almost every situation of physical relevance. As we
shall see, a bouncing Universe requires that it be violated in many
circumstances.

Perturbations of the stress energy tensor~(\ref{Tmunu}) read
\begin{equation} \delta T^0_0 = \delta\epsilon, \ \ \ \delta T^0_i
={\epsilon_0 +p_0\over a} \delta u_i, \ \ \ \delta T^i_j = -\delta p
\delta^u_j. \end{equation} Here, and in the rest of this paper (unless
otherwise stated), an index $0$ on a quantity is meant to be the
background, unperturbed value of this quantity.  These perturbations
are made gauge-invariant through the transformation $\delta \epsilon
\to \delta \epsilon ^{(\hbox{\sevrm gi})} = \delta  \epsilon +
\epsilon'_0 (B-E')$, $\delta p \to \delta p ^{(\hbox{\sevrm gi})} =
\delta p + p'_0  (B-E')$, and $\delta u_i \to \delta
u_i^{(\hbox{\sevrm gi})} = \delta u_i + a (B-E')_{;i}$. Einstein
equations then read, in terms of the Bardeen potential
\begin{equation} \nabla^2 \Phi - 3 \Hu \Phi' - 3 (\Hu^2 -\Ka) \Phi = 4
\pi G a^2 \delta \epsilon^{(\hbox{\sevrm
gi})},\label{EE1}\end{equation}
\begin{equation} (a \Phi)'_{;i} = 4 \pi G a^2 (\epsilon_0 +p_0) \delta
u_i^{(\hbox{\sevrm gi})},\label{EE2}\end{equation}
\begin{equation} \Phi'' + 3 \Hu \Phi' + (2 \Hu'+\Hu^2 -\Ka) \Phi = 4
\pi G a^2 \delta p^{(\hbox{\sevrm gi})},\label{EE3}\end{equation}
where the background equations ($\Hu \equiv a'/a$)
\begin{equation} \Hu^2 + \Ka = {8\over 3} \pi G a^2
\epsilon_0,\label{eps0} \end{equation}
\begin{equation} {2\over 3}\beta -\Hu^2 -\Ka = {8\over 3} \pi G a^2
p_0, \label{p0}\end{equation} have been taken into account. In
Eq.~(\ref{p0}), we have used the variable $\beta$ defined as
\begin{equation} \beta \equiv \Hu^2 -\Hu'+\Ka,\label{beta}
\end{equation}
which we will use intensively later on. It can be remarked, at this
stage, that Eqs.~(\ref{eps0}) and (\ref{p0}) imply
\begin{equation} \epsilon_0 + p_0 = {\beta \over 4\pi G
a^2},\end{equation} so that a violation of the Null Energy Condition
Eq.~(\ref{NEC}) will be equivalent to having $\beta < 0$.

In general, pressure perturbations can be separated into adiabatic and
entropic perturbations through
\begin{equation} \delta p = \left({\partial p\over\partial
\epsilon}\right) _S \delta \epsilon+ \left({\partial p\over\partial
S}\right) _\epsilon \delta S = \cs \delta \epsilon + \tau \delta
S,\label{dp}\end{equation} with $\delta S$ the entropy
perturbation. In many situations, $\cs$ is interpreted as the squared
sound velocity of the fluid and denoted $c_{_S}^2$. However, for  the
fluids which appear in  the examples below, this variable cannot be
interpreted as such. Already in plasma physics there are situations
where $\cs ^{1/2}$ is not the group velocity of the fluid (see
Ref. \cite{thorn}, page 406, for a discussion on this point, and
references therein). That is why we shall avoid calling $\sqrt{\cs}$
the sound velocity in what follows.

Using Eq.~(\ref{dp}), and plugging Eq.~(\ref{EE3}) into (\ref{EE1}),
one gets
\begin{eqnarray} \Phi'' + 3\Hu (1+\cs)\Phi' -  \cs \nabla^2 \Phi
& \hfill & \nonumber \\ + \left[ 2\Hu'+(1+3\cs)(\Hu^2 -\Ka)\right]\Phi
&=& 4\pi Ga^2 \tau \delta S,\label{I}\end{eqnarray} which becomes an
homogeneous equation in the case of adiabatic perturbations $\delta
S=0$ with which we shall be concerned from now on (see however the
discussion in Sec.~\ref{subsecB}).

We shall need, in what follows, two more pieces of information. It can
be seen, by combining Eqs.~(\ref{eps0}), (\ref{p0}) and (\ref{dp})
that the following useful identity for $\cs$
\begin{equation} \cs = {p_0'\over \epsilon_0'} = -{1\over 3} \left( 1+
{\beta'\over \Hu \beta}\right),\label{sound}\end{equation}
holds. Moreover, it is clear that perturbation theory can only make
sense if the perturbed Einstein tensor remains a small quantity. It
turns out that Einstein equations do not involve anything but the
Bardeen potentials. This stems from the fact that the gauge-invariant
part of the Einstein tensor $\delta G_\mu^{(\hbox{\sevrm gi})\nu}$
[the left-hand side of Eqs.~(\ref{EE1}) to (\ref{EE3})] only depends
on $\Phi$ and $\Psi$~\cite{mfb}. Satisfying the general requirements
for perturbative analysis thus demands that not only the Bardeen
potential, as discussed above, but also its first and second
derivative in time, in order to remain a perturbation, should be
regular and well-behaved for all times. As we shall see later, this is
not always a trivial requirement.

Hydrodynamical perturbations can be shown to originate from a single
gauge-invariant variable $v(\eta,{\bf x})$, whose dynamics is
derivable from the simple action
\begin{equation} S = {1\over 2} \int \sqrt{\gamma} \hbox{d}^4x \left(
v'^2 - \cs \gamma^{ij} v_{,i} v_{,j} + {z''\over z} v^2\right)
,\label{Sv}\end{equation} where $\gamma$ is the determinant of the
3-metric $\gamma_{ij}$ (with inverse $\gamma^{ij}$), and the
background function $z(\eta)$ is
\begin{equation} z^2 \equiv {a^2 |\beta|\over\Hu^2 |\cs|},\end{equation}
with $\beta$ defined in Eq.~(\ref{beta}). The variable $v$ is formed
with the fluid velocity perturbation potential and the Bardeen
potential, its explicit form being here however irrelevant.

Eq.~(\ref{Sv}) is in the correct form for a straightforward canonical
quantization of the field $v$. There exists a connection between $v$
and the Bardeen potential which we will use below and that we
therefore reproduce here. We first expand $\Phi$ on a basis of
eigenfunctions of the Laplace-Beltrami operator (here, $k$ is a
generic eigenvalue of this operator, which, depending on the curvature
$\Ka$, can be continuous or discrete),
\begin{equation} \Phi = \sqrt{{3\over 4}} \ell  {\beta^{1/2}\over
a}\int \hbox{d}k \left( \psi_k({\bf x}) u_k^* (\eta) a_k + \hbox{h.c.}
\right), \label{expansion} \end{equation} where $\ell \equiv\sqrt{8\pi
G/3}$ is the Planck length, and $\psi_k({\bf x})$ satisfies
\begin{equation} (\nabla^2 + k^2) \psi_k({\bf x}) = 0\; .\end{equation}
Implementing the usual commutation relations
\begin{equation} [a_k, a^\dagger_{k'} ] =\delta _{kk'},\end{equation}
(the $\delta$ here being a Dirac distribution or a Kronecker symbol
depending on whether the space is flat or open, or closed,
respectively) yields the time evolution equation for the
wavelength-dependent modes of the Bardeen potential
\begin{equation} u''_k(\eta) + \left( \cs k^2 - {\Theta''\over \Theta}
\right) u_k(\eta) =0, \ \ \ \Theta \equiv {1\over \sqrt{|\cs|}
z}. \label{uk}\end{equation} An expansion similar to
Eq.~(\ref{expansion}) can be done for the field $v$. It turns out that
its modes $v_k(\eta)$ satisfy
\begin{equation} v''_k(\eta) + \left( \cs k^2 - {z''\over z}
\right) v_k(\eta)=0, \label{v} \end{equation} and they are related to
the Bardeen operator through the relation\footnote{When $\beta$ or
$\cs$ change sign, it is necessary to calculate $\Phi$ directly by
means of Eq.~(\ref{I}).}
\begin{equation} \Phi _k(\eta) \propto {\sqrt{|\beta|}\over |\cs|}
{z\over a} \left( {v_k\over z}\right)'.\label{uv} \end{equation} A
last interesting point concerning this field is that the eigenmodes
are expected to be normalized as
\begin{equation} v_k'(\eta) v^*(\eta) - v^{*\prime}_k(\eta) v_k(\eta) =
2 i,\end{equation} for all times $\eta$.

The process of generating primordial fluctuations then goes on by
requiring the Universe to be empty at some early time, i.e. to be in
the vacuum state $|0\rangle$ for which $a_k |0\rangle = 0$ for all
wavenumber $k$. The evolution of $v$ through Eq.~(\ref{v}) then
necessarily changes the state and dynamically produces particles that
are subsequently interpreted as primordial perturbations.

The background equations presented in this section will now be used
for the forthcoming analysis of the bouncing background, to which we
now turn.

\section{General bouncing behavior}
\label{sec:bounce}

In this section we will enumerate some properties of a background
spacetime with a bounce, and calculate some relevant quantities around
the bounce which will be important for the calculations of the
perturbations. We shall be mostly concerned in what follows by actual
bounces, but the pseudo-bounce situation is also included in our description.

Our sole assumption on the background is that, near the last bounce
before the standard model evolution took place, $a(\eta)$ can be
written as follows
\begin{equation}
\label{abounce}
a=a_0 + b\eta ^{2n} + d\eta ^{2n+1} + e\eta ^{2n+2}\; ,
\end{equation}
where the integer\footnote{Or half-integer in the case of a
pseudo-bounce.} $n$ satisfies $n\geq 1$. It means that $a(\eta)$ must
be at least ${\rm C}^{2n+2}$ near the bounce. In order that
Eq.~(\ref{abounce}) indeed represents a bounce, the otherwise
arbitrary parameter $b>0$. Our results are independent on what happens
before this last bounce (other bounces, a PBB evolution, or even a
singularity).  In the case where $\epsilon_0=\epsilon_0(a)$, which is
possible if and only if $a(\eta)$ is even, the presence of a bounce
implies that the model is free of singularities. This is the case of
models of Refs. \cite{seventies,bounce}.

For this model to be realistic, we require $a_0$ to be less (or even
much less) than $a_n \sim 10^{18}$~cm, where $a_n$ is the value of the
scale factor in the beginning of nucleosynthesis.

The function $\Hu (\eta)$ coming from Eq.~(\ref{abounce}) reads
\begin{eqnarray}
\label{Hbounce}
\Hu&=&\frac{1}{a_0^2} [2nba_0\eta ^{2n-1} + (2n+1)a_0
d\eta^{2n}\nonumber \\ & &+(2n+2)ea_0\eta ^{2n+1} - 2nb^2 \eta^{4n-1}],
\end{eqnarray}
while $\beta(\eta)$ is
\begin{eqnarray}
\label{bbounce}
\beta&=&\frac{1}{a_0^2}[\Ka a_0^2 -a_0 2n(2n+1)d\eta ^{2n-1}-
a_02n(2n-1)b\eta ^{2n-2} \nonumber\\ & & -a_0(2n+2)(2n+1)e\eta ^{2n} +
2n(6n-1)b^2\eta ^{4n-2}].
\end{eqnarray}
{}From these functions we can calculate the coefficient $\cs$ given in
Eq.~(\ref{sound}), yielding~:

\begin{eqnarray}
\label{csbounce}
\cs &=&-\frac{a_0}{6 n b}\biggl\{\frac{-a_0 2n(2n-1)(2n-2)b} {\eta ^2}
\nonumber\\ & & + 2 [\Ka b - (2n+2)(2n+1)e]a_0 n \nonumber \\ & & -
{2a_0 n(4n^2-1)d\over \eta} \nonumber\\ & & + 16b^2 n^2 (2n-1)\eta
^{2n-2}  + ... \biggr\} \nonumber\\ & & (\Ka a_0^2 -a_0 2n(2n-1)b\eta
^{2n-2}+ ...)^{-1}\; .
\end{eqnarray}

We will now evaluate the behavior of the background functions
$\Hu,\beta ,\cs$ and $z$ near the bounce for the possible values the
free parameters can assume.

\begin{enumerate}

\item\underline{$n>1$ and $\Ka \neq 0$.}

In this case, we find the following behaviors
\begin{eqnarray}
\cs &=& \frac{(2n-1)(2n-2)}{3\Ka \eta ^2}, \label{cs21} \\ \Hu &=&
\frac{2nb}{a_0}\eta ^{2n-1}, \label{H1} \\ \beta &=&\Ka, \label{beta1}
\\ z&\propto& \frac{1}{\eta ^{2n-2}}\label{z1} \; ,
\end{eqnarray}
so that it turns out to be possible to have a bounce without violation
of the NEC only if $\Ka = 1$. Note that this case, as well as the
following one, can accommodate a pseudo-bounce (where $2n$ is an odd
number strictly greater than $1$).

\item\underline{$n>1$ and $\Ka = 0$.}

For the special case of a flat background, the various quantities
needed to describe perturbations are modified as
\begin{eqnarray}
\cs &=& - \frac{a_0(2n-2)}{6bn \eta ^{2n}},\label{cs22} \\ \Hu &=&
\frac{2nb}{a_0}\eta ^{2n-1},\label{H2} \\ \beta &=&
-\frac{b}{a_0}2n(2n-1)\eta ^{2n-2},\label{beta2} \\ z &\propto& {\rm
const.}\label{z2}\;
\end{eqnarray}

Here there is always violation of the NEC as $\beta\leq 0$ at least in
some open neighborhood of the bounce.

\item\underline{$n=1$, $d\neq 0$, $\forall\Ka$.}

This is the case where the second derivative of $a(\eta)$ is non
vanishing and $a(\eta)$ is not even. The relevant quantities are

\begin{eqnarray}
\cs &=& -\frac{a_0d}{b (2b-\Ka a_0)\eta},\label{cs23} \\ \Hu &=&
\frac{2b}{a_0}\eta,\label{H3} \\ \beta &=&\Ka -
\frac{2b}{a_0},\label{beta3} \\ z &\propto& \frac{1}{\sqrt{\eta}}
\label{z3}\; .
\end{eqnarray}

In this case, we can also have a bounce without violation of the NEC
if $\Ka = 1$ and $2b/a_0 \leq 1$.  However, this condition is not
physically plausible, as it is argued in the Appendix.

\item\underline{$n=1$, $\forall \Ka$, and $d=0$.}

Finally, for the case that can represent an even $a(\eta)$, we get
\begin{eqnarray}
\cs &=& \frac{8b^2+(\Ka b - 12 e)a_0}{3b (2b-\Ka a_0)},\label{cs24} \\
\Hu &=& \frac{2b}{a_0}\eta,\label{H4} \\ \beta &=&\Ka -
\frac{2b}{a_0},\label{beta4} \\ z &\propto& \frac{1}{\eta}
\label{z4}\; .
\end{eqnarray}

Here again, we can have a bounce without violation of the NEC if $\Ka
= 1$ and $2b/a_0 < 1$.

This case corresponds to models of Refs.~\cite{seventies,bounce}.

\end{enumerate}

\medskip

Note that there are exceptional cases when $\Ka = 1$ and $2b=a_0$ in
cases 3 and 4, and when $8b^2+(\Ka b - 12 e)a_0 = 0$ in case 4.  For
case 3 with $\Ka = 1$ and $2b=a_0$, it can be shown that unbounded
growth occurs in the bounce, but we will not exhibit these
calculations below because, as we have already mentioned, $\Ka = 1$
and $2b=a_0$ are not together plausible in realistic models. For case
4, we will show in the next section that the crucial instant of
instability is not in the bounce itself, but when the Universe makes
the transition from the region where the NEC is violated to the region
where it is not. The occurrence of this transition is independent on
the value of $8b^2+(\Ka b - 12 e)a_0$.

\section{Non-linear collapse of non singular models}
\label{sec:pert}

Having set the general conditions under which a bounce is expected to
occur in the early Universe, we now apply the cosmological
perturbation theory recalled in Sec.~\ref{sec:rappels} to such a
background. In particular, we shall make extensive use of
Eqs.~(\ref{I}) and (\ref{uk}--\ref{uv}) to derive the leading terms in
the behavior of the gauge-invariant Bardeen perturbation potential.
Let us turn back to the time the bounce is supposed to have taken
place, and consider the various situations for the free parameters
already discussed above.

\subsection{The bounce}

There are four distinct cases near the bounce which deserve to be
examined.

\begin{enumerate}

\item\underline{$n>1$ and $\Ka \neq 0$.}

The first case is $n>1$ with non vanishing curvature $\Ka =\pm
1$. Eqs.~(\ref{cs21}) to (\ref{z1}) transform Eq.~(\ref{v}) into
\begin{equation} v_k'' + {(2n-1)(2n-2)\over\eta^2}\left( {k^2\over 3\Ka}
-1 \right) v_k =0, \end{equation} whose solution is $v= v_+ \eta^{p_+}
+ v_- \eta^{p_-}$ ($v_+(k)$ and $v_-(k)$ being two integration
constants), with $2 p_\pm \equiv 1\pm
\sqrt{1-4(2n-1)(2n-2)\biggl({k^2\over 3\Ka}-1\biggr)}$.  For negative
curvature space, $\Ka =-1$, one has $p_-<0$, for whatever value of
$k$, while for $\Ka =1$, $p_-<0$ provided $k^2<3$. Note that $\forall
k$, $(p_- -1)$ has a negative real part in both spaces $\Ka=\pm 1$.

Inserting this form into Eq.~(\ref{uv}) yields, for the Bardeen
potential, $\Phi_k \propto \eta^{p_-}$, i.e. also a divergent quantity
for $p_- <0$. It can be seen from the previous considerations that,
irrespective of the value of the wavenumber $k$, there will be a
divergent quantity either in the metric perturbation directly for
$\Ka=-1$, or in its derivative ($\propto \eta^{p_- -1}$) for $\Ka=1$.

\item\underline{$n>1$ and $\Ka = 0$.}

The second case, similar to the first ($n>1$) but for flat spatial
sections ($\Ka =0$), has $z=\hbox{const}$. In order to calculate , the
``pump'' term, we need to obtain $z$ in the next to leading order of
approximation. However, making use of the expansion formula for $v_k$

\begin{eqnarray}
v_k &=& z\biggl[A_1 + A_2\int ^{\eta}_{\eta
_0}{\hbox{d}\hat{\eta}\over {\hat{z}}^2} \nonumber \\ & &- k^2 \int
^{\eta}_{\eta _0}{\hbox{d}\hat{\eta}\over {\hat{z}}^2}\int
^{\hat{\eta}}_{\eta _0}\hbox{d}\tilde{\eta}{\widetilde{\cs}}
{\tilde{z}}^2\biggl(A_1 + A_2\int ^{\tilde{\eta}}_{\eta
_0}{\hbox{d}\tilde{\tilde{\eta}}\over{\tilde{\tilde{z}}}^2}+\cdots
\biggr], \nonumber
\end{eqnarray}
one can verify that $v_k$ contains a term proportional
$k^2/\eta^{2n-2}$, which diverges in the bounce.

In order to calculate what happens to the Bardeen potential $\Phi _k$,
it is easier to investigate directly the Bardeen equation (\ref{I})
when these values of the parameters are assumed, and one finds
\begin{equation} \Phi_k''+{2-2n\over\eta} \Phi_k' - {a_0 (n-1) k^2\over
3 b n \eta^{2 n}}\Phi_k =0,\end{equation} whose solution, setting

\begin{equation}
\alpha \equiv k\sqrt{a_0 / [3 n b (n-1)]}\eta^{1-n}\; , \nonumber
\end{equation}
is given by (see Eq.~(8.491/3) in Ref.~\cite{Grad})
\begin{equation} \Phi_k = \eta^{(2n-1)/2} \left\{ \Phi_+ I_
{ {2n-1\over 2n-2}} (\alpha ) + \Phi_- K_{ -{2n-1\over 2n-2}}
(\alpha) \right\},
\end{equation}
with two arbitrary constants of integration (depending on $k$)
$\Phi_+$ and $\Phi_-$. In this paper the symbols $J,N$ and $I,K$ stands
for pairs of linearly independent solutions of the Bessel equation, Bessel
and modified Bessel functions of first and second kind. As, by assumption,
$n>1$, we
find that the leading order behavior of the Bardeen potential is
[Ref.~\cite{Grad}, Eqs.(8.451/3-4)]
\begin{equation} \Phi_k \sim \eta^{(3n-2)/2} \hbox{e}^{|\alpha|},
\end{equation}
namely, this perturbation exhibits a singular behavior near the
bounce ($|\alpha | \to \infty$).

\item\underline{$n=1$, $d\neq 0$, $\forall \Ka$.}

This is the most general case, where $a(\eta)$ is not even and its
second derivative is not null. Eqs.~(\ref{cs24}--\ref{z4}) when
inserted into Eq.~(\ref{v}), yield
\begin{equation} v_k'' + \biggl({\lambda (k)\over \eta}  -
{3\over 4\eta^2}\biggr) v_k =0.
\end{equation}
The explicit form of $\lambda$ is not important. The solution of this
equation reads [Ref.~\cite{Grad}, Eqs.(8.404/1-2)]
\begin{equation} v_k = \sqrt{\eta} \left\{v_+ J_2
(2\sqrt{\lambda\eta}) + v_- N_2 (2\sqrt{\lambda\eta})\right\},
\end{equation}
where
$v_+,v_-$ are arbitrary constants depending on $k$. The leading order
behavior of the divergent term is $v_k\propto 1/\sqrt{\eta}$.  The
Bardeen potential can be obtained from Eq.~(\ref{uv}), or directly
from Eq.~(\ref{I}) (with $\delta S =0$), yielding, near $\eta =0$,
\begin{equation} \Phi _k \propto \sqrt{\eta}\left\{v_+ J_1
(2\sqrt{\lambda\eta}) + v_- N_1 (2\sqrt{\lambda\eta})\right\},
\end{equation}
which is convergent at $\eta =0$. However, the first derivative of
$\Phi _k$ has a divergent term,
\begin{equation} \Phi_k ' \propto \sqrt{\eta}N_0
(2\sqrt{\lambda\eta}),
\end{equation}
which behaves as $\ln \eta$ near $\eta =0$, and the second derivative is
\begin{equation} \Phi_k '' \propto {1\over\eta},
\end{equation}
yielding unbounded growth of the perturbed Einstein tensor (curvature)
near the bounce.

\item\underline{$n=1$, $\forall \Ka$, and $d=0$.}

This is the case which contains models from
Refs.~\cite{seventies,bounce}, namely, that for which the first term
in the scale factor expansion is quadratic ($n=1$), for whatever value
of the spatial curvature $\Ka$, and which can accommodate the case for
which $a(\eta)$ is even.

Eqs.~(\ref{cs23}--\ref{z3}), inserted into Eq.~(\ref{v}), yield
\begin{equation} v_k'' - {2\over \eta^2} v_k =0,
\end{equation}
yielding the divergent mode $v_k\propto 1/\eta$. For the Bardeen
potential, we can use Eq.  (\ref{uk}), or directly Eq. (\ref{I}) (with
$\delta S =0$), to obtain that, near $\eta =0$, it behaves as

\begin{equation} \Phi_k \propto \cos(\alpha(k,\Ka)\eta)\;
,\end{equation} which is regular up to second order derivatives
independently on the form of the function $\alpha(k,\Ka)$.

This is a case where there is no divergence in the bounce
itself. However, as shown in the Appendix, near the bounce of a
realistic Universe of this type, $\beta < 0$, even when $\Ka =1$. As
the Universe must have reached a region in which the NEC must be
satisfied at later times, there must exist an instant $\eta_0$ at
which the function $\beta$ vanishes. As it turns out, this is a
critical point also from the point of view of perturbation theory.

\end{enumerate}

\subsection{NEC transition}

In what follows, we now concentrate on the point where $\beta=0$, so
that we shift the origin of time~: for now on, $\eta_0=0$ for
$\beta_0$, and we denote by an index $0$ quantities evaluated at this
point.

We now assume that the scale factor around $\eta=0$ is, again,
derivable at least up to third order, so that the following expansion
\begin{equation} a(\eta) = a_0 \left[ 1+\Hu_0\eta +{1\over 2}
(2\Hu_0^2 +\Ka) \eta^2 + {1\over 3 !} a_3 \eta^3 + \cdots \right],
\label{a0}\end{equation}
holds. In this relation, $a_3 \equiv a'''(0) /a_0$, and use has been
made of
\begin{equation} {a''(0)\over a_0} = 2\Hu_0^2+\Ka,\end{equation}
which is a simple rewriting of $\beta_0 = 0$.

Using the expansion~(\ref{a0}), we find that
\begin{eqnarray} \Hu &=& \Hu_0+\eta (\Hu_0^2+\Ka)\nonumber \\
& & +\eta^2 ({a_3\over 2} -2\Hu_0^3-{3\over 2} \Ka \Hu_0) +{\cal
O}(\eta^3),\end{eqnarray} and
\begin{equation} \beta = (6\Hu_0^3+5 \Ka \Hu_0 -a_3) \eta +{\cal
O}(\eta^2).\end{equation} As shown in the Appendix, $\beta _0 '\neq 0$
for realistic models and hence $6\Hu_0^3+5 \Ka \Hu_0 -a_3 \neq 0$. As
a result,
\begin{equation} \cs = - {1\over 3 \Hu_0 \eta}
+{\cal O} (\eta^0),\end{equation} which transforms Eq.~(\ref{I}) near
$\eta =0$ into
\begin{equation} {d^2\Phi_k\over d\eta ^2} + {1\over\eta}{d\Phi_k
\over d\eta}+\biggl({\Ka\over\Hu_0} -\Hu_0 -{k^2\over
3\Hu_0}\biggr){\Phi_k\over \eta} =0,\label{65}\end{equation} whose
solution reads~(\cite{Grad}, Eq.(8.494/5))
\begin{eqnarray} \Phi_k &=& \Phi_{_I} \eta I_2
\left[2\sqrt{(-{\Ka\over\Hu_0} +\Hu_0 +{k^2\over 3\Hu_0})
\eta}\right]\nonumber \\ & & + \Phi_{_K} \eta K_2
\left[2\sqrt{(-{\Ka\over\Hu_0} +\Hu_0 +{k^2\over 3\Hu_0})
\eta}\right],\end{eqnarray} where $\Phi_{_I}$ and $\Phi_{_K}$ are
arbitrary constants depending on $k$, while $I_2$ and $K_2$ are
ordinary Bessel functions. For realistic models, $\Hu_0 \gg 1$.  To
lowest order, the Bardeen potential is then found as
\begin{equation} \Phi_k \sim \hbox{const.} + {\cal O}
(\eta),\end{equation} which would seem to imply a safe perturbative
expansion. However, its second derivative with respect to conformal
time exhibits, irrespective of the value of $k$, the mildly diverging
behavior
\begin{equation} \Phi_k''\sim \ln \eta \end{equation}
so that, by virtue of Eqs.~(\ref{EE1})~--~(\ref{EE3}), we are forced
to conclude that even this perturbative expansion generates unbounded
growth.

\section{Discussion}
\label{sec:discuss}

The various cases that have been investigated here represent an
exhaustive list of all the possibilities that satisfy the requirements
of {\it i)} having a non-strictly monotonic behavior for the scale
factor at some instant of time, {\it ii)} that this instant of time be
in our past light cone, and that {\it iii)} therefore, the Universe in
which the bounce took place is ours, in the sense that nucleosynthesis
as we know it occurred. In all these cases, a divergent behavior for
the cosmological hydrodynamical perturbations have been found, which are
responsible
for an observationally excluded non-linear growth of primordial
inhomogeneities. We would like to discuss three points regarding these
results.

\subsection{$\cs\to\infty$.}
\label{subsecA}

The first important point we should like to emphasize concerns the
divergent behavior for $\cs$, which could be expected to be
responsible, right from the  outset, for the divergence of the
perturbations. That this  is not the case can be seen on the following
example, based on the last case of the previous section. If $\beta
'=0$, i.e., if $\beta \propto \eta ^{2n+1}$ with $n \geq 1$, when
$\beta =0$, then, to leading order, $\cs$ behaves as $\cs \sim -
{2n+1\over 3 \Hu_0 \eta}$, and Eq.~(\ref{65}) is transformed in such a
way that the Bardeen potential now reads
\begin{eqnarray} \Phi_k &=& \Phi_{_I} \eta ^{n+1} I_{2n+2}
\left[2\sqrt{(-{\Ka\over\Hu_0} +\Hu_0 +{k^2\over 3\Hu_0})
\eta}\right]\nonumber \\ & & + \Phi_{_K} \eta ^{n+1} K_{2n+2}
\left[2\sqrt{(-{\Ka\over\Hu_0} +\Hu_0 +{k^2\over 3\Hu_0})
\eta}\right],\end{eqnarray}
which has convergent derivatives up to order $2n+2$. This is a case
where, even in the point where $\cs$ diverges, the curvature and all
other relevant quantities converge. It happens because $\delta
\epsilon$ goes to zero at this point. Hence, stable models with a
divergent $\cs$ at one point are indeed possible, as this simple
example testifies. It is the combination of all the conditions {\it
i)}, {\it ii)} and {\it iii)}, that gives rise to unbounded growth.

\subsection{Gauge invariance}
\label{subsecB}

In each of the cases investigated above, we have found a growing mode
for the Bardeen potential, which leads to divergences either in the
metric perturbation or in the curvature. In view of recent discussions
of similar modes found in the PBB case for instance~\cite{brustein}, a
natural question to ask then is whether they are actually
physical. Here, we should like to argue that indeed, and contrary to
their counterpart in PBB, they cannot be ``gauged-down''. In this
sense, they represent real physical inhomogeneities, doomed to grow
non-linear.

It is well-known that for wavelength larger than $\Hu^{-1}$, a
residual gauge artifact might exist that can be tamed for $k\ll \Hu$
(for instance by going to an off-diagonal gauge in which higher order
derivatives  are involved~\cite{brustein}). However, the divergences
we have obtained in the last section are present for all
wavelengths. Furthermore, contrary to the PBB case, our divergences
occurs on regular points of spacetime, where $\cal{H}$ is regular
(even zero in the bounce, implying that all wavelengths are smaller
then the Hubble radius at this time), and it is not possible to blame
a bad behavior of $\cal{H}$ for their existence.

Let us see that explicitly. Assume one makes the calculations in the
longitudinal gauge for which $B=E=0$ (note that this is not the case
of the present paper). In this gauge, Eq.~(\ref{dg}) simplifies to
\begin{equation} \hbox{d}s^2 = a^2(\eta) \left[ (1+2 \phi
)\hbox{d}\eta^2 -(1-2\psi)\gamma_{ij} \hbox{d}x^i \hbox{d}x^j
\right],\label{dg2}\end{equation} where both $\psi$ and $\phi$ take
the gauge-invariant value $\Phi$. We now assume that there exists a
gauge transformation $x^\alpha \rightarrow \tilde x^\alpha = x^\alpha
+ \xi^\alpha ({\bf x},\eta)$ that could cancel the divergence in the
potential $\Phi_k$, leading to (in Fourier space, $\xi_k^\alpha
(\eta)$ being the $\Ka-$dependent Fourier transform of the
infinitesimal vector field $\xi^\alpha$) $\tilde\phi_k = \Phi_k -\Hu
\xi_k^0 (\eta) -\xi_k^{0\prime}(\eta)$, $\tilde \psi_k = \Phi_k + \Hu
\xi_k^0$, $\tilde B_k = \xi_k^0 - \xi_k'$ and $\tilde E_k =
-\xi_k$. The second of these relations reveals that since $\Phi_k$ is
a divergent quantity at the physical point of space-time under
consideration, the relevant transformation must be singular in order
to produce a bounded value for $\psi_k$. This is to be contrasted with
the PBB case for which, as $\eta_{_{PBB}}\to 0$, $\Hu_{_{PBB}} \to
\eta^{-1}$ so that the gauge vector $\xi$ can indeed assume
infinitesimal values while canceling the divergence. In our case,
$\Hu\to\Hu_0$, a constant which, at  the bounce, actually vanishes,
thereby rendering the transformation even more singular. However we
look at these perturbations, they do grow infinitely, leading to
singularities, or at least to large (non-linear) inhomogeneities.

This can also be seen by using the gauge invariant curvature
perturbation variable~\cite{curvar}
\begin{equation} \zeta = \Phi +{\Hu^2 + \Ka\over \Hu\beta} (\Phi' +
\Hu \Phi),\end{equation} which also exhibits divergences in the cases
of interest here as can readily be checked. However, one should use
this variable cautiously as the corresponding divergences may not be
physical~: in deriving this form for the three-curvature modes, it is
explicitly assumed that the NEC is satisfied, so that for $\beta\to
0$, a fake divergence can occur.  This is indeed what happens in the
situation for which $\beta\propto \eta^{2n+1}$ mentioned in
Sec.~\ref{subsecA}~: in this case, $\zeta\to\infty$ even though
nothing particular happens in the theory at this point.

\subsection{Adiabatic and entropy perturbations}
\label{subsecC}

The analysis that is presented here is restricted to purely adiabatic
perturbations, which might be seen as too restrictive  an hypothesis
in view of the fact that at least two barotropic fluids are needed in
a hydrodynamical description to yield a bounce. Bardeen
equation~(\ref{I}) for  the potential $\Phi$ can be read as an
inhomogeneous equation with a source term proportional to the entropy
perturbation, if the latter is independent of the former. In this
situation, its solution is given by a particular solution  of the full
inhomogeneous equation, whatever it may be, plus the general solution
of the homogeneous one, i.e. Eq~(\ref{I}) with $\delta S=0$, which is
the equation we have been discussing throughout.  As it was found that
the adiabatic perturbations do exhibit a singular behavior near the
bounce or at the time where the NEC is recovered, all the
perturbations will exhibit such a behavior.

What if the entropy perturbation depends on the Bardeen potential~? In
that case, it is useful to recall that an arbitrary perturbation can
be decomposed over its possible modes (in the linear regime with which
we are concerned), namely one adiabatic, and $N-1$ isocurvature modes
for the case of $N$ constituents. All these modes may be well behaved,
but one sees that with the decomposition~\cite{riaz}
\begin{equation} \Phi = \Phi_{\hbox{\sevrm adiabatic}} +
\sum_{i=1}^{N-1} \Phi^{(i)}_{\hbox{\sevrm
isocurvature}},\end{equation} it is sufficient that the adiabatic mode
we have been studying diverges to yield a cosmological catastrophe,
unless the isocurvature modes somehow compensate exactly this
divergence. Here, again, we return to recent observations of the
CMB~\cite{CMBdata} according to which the dominant part of the
primordial perturbations was in the adiabatic mode. Therefore, once
again, the requirement that the bouncing Universe be ours implies a
stringent constraint~: in this case, it is that the adiabatic mode
under scrutiny here is not negligible, and therefore cannot be
compensated by an isocurvature mode.

\section{Conclusions}
\label{sec:conc} 

The simplest way to envisage a cosmological model without an initial
singularity is to think that the scale factor never shrunk to zero but
reached a minimal value, preceded by a contraction phase.  In fact,
much of non-singular cosmological models proposed in the literature
present this bouncing behavior.

We have shown in the present paper that any bouncing solution whose
matter content is described by hydrodynamical fluids is
highly unstable, at least if they are to be smoothly joined with our
observed Universe.  In particular, applying to these solutions the now
standard cosmological hydrodynamical perturbation theory, one finds that
in all cases
of interest, the Bardeen potential, which represents the typical
scalar metric fluctuations, possesses divergent modes near the bounce
that cannot be gauged away. For the quantum modes, these divergences
appear because the so-called ``pump'' term $z''/z$ acts as an
infinitely deep potential well~: the corresponding $v-$wave then turns
out to be unbounded for $\eta =0$, so that an infinite amount of
particles are produced.  This is similar to the divergent mode in
ordinary inflationary scenarios, with one fundamental difference~: in
the usual scenarios, the divergence comes from the singular point
($a=0$) of the metric, and can therefore safely be neglected, or
altogether ignored, on the basis that physics cannot be described by
any known theory at this point.  Near a bounce, however, such an
argument fails as one may construct perfectly consistent theories
according to which Einstein gravity and hydrodynamics should hold at
the bounce.

The divergent behavior which is obtained {\sl in all relevant cases}
actually means that any perturbation present at the time a bounce
occurs would rapidly become non-linear. In some cases, this unbounded
growth happens in the bounce itself.  If, in addition, the bounce
implies a violation of the NEC, non linearities appear in at least one
other situation~: when the NEC is recovered. Hence, in many models,
one have divergences in two or more moments of their  history. As such
a phenomenon happens for all wavelengths (the Hubble radius begin
infinite near the bounce), one would then expect an ensemble of very
strong inhomogeneities to be formed prior to any regular phase of the
Universe.\footnote{There is the very specific case where the initial
conditions, for all wavelength $k$, are chosen such that the diverging
mode is vanishing at all times, and if all the solutions of the
inhomogeneous equation, when $\delta S \neq 0$, are regular.  Such an
extreme choice cannot, of course, be ruled out on purely mathematical
grounds, but it seems fairly unlikely.  In the models where the
divergences appear more then once in their history, even
mathematically this may not be possible.}

Note that this does not preclude the actual Universe to have a remote
bounce. It is possible that after the inhomogeneities are formed, a
small region of the Universe begins to inflate and eventually yields
our Universe, much like in chaotic inflation, rendering the bounce
spacelike separated from us (thereby departing from our hypothesis of
physically observable bounce). The bounce in itself is very likely to
be quite far from any observational verification. Its only interest,
at the present time, is to reveal that it is theoretically  possible
to construct non singular cosmological models.

Summarizing, a Universe filled only with hydrodynamical fluids
and with an observable bounce would already have
been very highly inhomogeneous by the time nucleosynthesis
initiates. This conclusion is, from the point of view of observational
cosmology, inadmissible. This shows that one at least of the following
statement is true~:

\medskip \noindent i) No observable bounce ever took place in the
Universe, or, stated differently, the Universe always expanded.

\noindent ii) If an observable bounce took place, then the dominant
contribution in the stress-energy sourcing Einstein equations was not
expressible in the form of a perfect fluid so that hydrodynamic
cosmological perturbations do not make sense.

\noindent iii) Einstein equations are not valid in the time region
where the Universe bounced.

\medskip

Hence, in the framework of our hypothesis (4-dimensional GR and
hydrodynamical
perturbations), we are therefore led to postulate a {\sl No-Bounce
Conjecture} according to which the opening question can be positively
answered~: from a pragmatic point of view, the Universe has always
expanded.

\acknowledgments

We would like to thank CNPq of Brazil for financial support.  PP
should like to acknowledge CBPF for hospitality during the time this
work was being done. We also would like to thank N.~Deruelle,
J.~Martin and the group of ``Pequeno Semin\'ario'' for various
enlightening discussions.

\section*{APPENDIX~: Connecting the bounce with the radiation dominated
era of the standard model}

In this appendix we will prove that $\beta < 0$ and $\beta ' \neq 0$
near the bounce when the conditions of case 4 are valid ($a(\eta) =
a_0 + b\eta ^2 + ... $ in the neighborhood of the bounce).

\subsection{}

To prove that $\beta < 0$, we will restrict ourselves to the single
non trivial case $\Ka =1$, where $\beta = 1-2b/a_0$.

Considering conformal time ranging from a neighborhood of the last
bounce until standard model evolution ($a$ has the minimum value $a_0$
in this range), one can define without ambiguity the function

\begin{equation}
\label{f}
f(\eta) = \sqrt{a^2(\eta) - a_0^2\over 2ba_0} \; ,
\end{equation}
or the inverse relation

\begin{equation}
\label{af}
a(\eta) = \sqrt{a_0^2 + 2 b a_0 f^2(\eta)} \; ,
\end{equation}
where all initial conditions and physical parameters of the solution
are contained in $b$ and $a_0$, which means that $f(\eta)$ is a pure
a-dimensional mathematical function.

Near the bounce we obtain that $f(\eta)=\eta + O(\eta^3)$ (remember
that this is the case $d=0$).  Furthermore, any such model must be
connected with the cosmological standard model before nucleosynthesis,
which means that from $\eta =\eta_n \sim 10^{-11}$ to $\eta =
\eta_{eq} \sim 10^{-4}$ (equal matter and radiation equilibrium) one
should have
\begin{equation}
a(\eta) \sim \bar{a} \sin (\eta) \sim \bar{a} \eta \; ,
\end{equation}
where $\bar{a}\sim 10^{29}$cm.  Substituting this expression in
Eq. (\ref{f}), and knowing that $a_0 < a_n \sim 10^{18}$ cm, we get
\begin{equation}
\label{f2}
f(\eta) \sim \frac{\bar{a}}{\sqrt{2ba_0}} \eta \; .
\end{equation}

However, we know that around $\eta = 0$ the function $f$ is exactly

\begin{equation}
\label{f3}
f(\eta) = \eta + {f'''(0)\over 6!} \eta ^3 + ... \; .
\end{equation}
If $2b/a_0 < 1$ then $2ba_0<a_0^2<a_n^2$ and
$\bar{a}/\sqrt{2ba_0}>10^{11}$. Knowing that Eq. (\ref{f2}) is valid
already at $\eta \sim 10^{-11}$, the compatibility of Eqs. (\ref{f2}),
(\ref{f3}) and $2b/a_0 < 1$ is possible only if $f'''(0)$ and higher
derivatives be a-dimensional huge number, which must be extremely fine
tuned in order to yield the factor $1$ multiplying the $\eta$ term in
the above expansion.\footnote{One possible mathematical function which
is compatible with these requirements is $f(\eta) = (1-10^{14})\eta
\exp (-10^{25}\eta ^2) + 10^{14}\eta $.  Note, however, the huge
numbers present in this function, and the presence of the term
$1-10^{14}$. When $\eta$ goes to zero, the difference of two numbers
of order $10^{14}$ must be exactly $1$, an extreme fine
tuning~!} Physicists believe that no reasonable dimensionless
function $f(\eta)$ should contain such fine tuned very large
a-dimensional numbers, so we conclude that $f(\eta)= \eta + ...$ for
$\eta < \eta _{eq}\sim 10^{-4}$\footnote{One conclusion from these
considerations is that the scale factor solution obtained in Ref.
\cite{bounce} is the most general physically reasonable one, up to
matter domination, which can accommodate a bounce with our real
Universe.}, and that $\bar{a}=\sqrt{2ba_0}\sim 10^{29}$cm which
implies that $2b/a_0 > 10^{22}$. Hence, $\beta = 1 - \frac{2b}{a_0}$
is negative, and also in this case the NEC is violated.

\subsection{}

We will now prove that $\beta '\neq 0$ around the point where $\beta$
changes sign. The change in sign of $\beta$ must occur before
nucleosynthesis. Hence we can write
\begin{equation}
\label{a2}
a(\eta) = \sqrt{2ba_0 \eta ^2 + a_0^2} \; .
\end{equation}
Calculating $\beta$ given in Eq. (\ref{beta}) and equating it to zero
we get the following quartic equation~:

\begin{equation}
\label{roots}
\Ka 4a_0^2b^2\eta ^4 + 4ba_0^2(\Ka a_0 + 2b)\eta ^2 + a_0^3(\Ka a_0 -
2b) = 0 \; .
\end{equation}
If $\beta '=0$ when $\beta = 0$, then $\beta ''$ must also be zero.
If it is not, then $\beta$ is not changing sign. Hence, $\beta$ must
be of the form $\beta = A(\eta-\eta_1) (\eta -\eta_0) ^3$ if $\beta
'=0$, where $\eta_0$ is some root of the quartic equation
(\ref{roots}), $\eta_1\not= \eta_0$, and $A$ is a constant. But the
polynomial form of Eq.~(\ref{roots}) does not contain any cubic
$\eta^3$ term so that $\beta$ cannot have three equal roots. Hence, we
must conclude that $\beta '\neq 0$.

\end{document}